\documentclass[prl,reprint,amsmath]{revtex4-1}
\usepackage{amsfonts}
\usepackage{color}
\usepackage[colorlinks]{hyperref}
\hypersetup{urlcolor=blue}
\begin{document}
\title{Sequential measurement of conjugate variables as an alternative quantum state tomography}
\author{Antonio \surname{Di Lorenzo}}
\affiliation{Instituto de F\'{\i}sica, Universidade Federal de Uberl\^{a}ndia,\\
 38400-902 Uberl\^{a}ndia, Minas Gerais, Brazil}
\begin{abstract}
It is shown how it is possible to reconstruct the initial state of a
one-dimensional system by measuring sequentially two conjugate variables. The procedure relies on the  quasi-characteristic function, the Fourier-transform 
of the Wigner quasi-probability. The proper characteristic function obtained by Fourier-transforming the experimentally 
accessible joint probability of observing ``position" then ``momentum" (or \emph{vice versa}) 
can be expressed as a product of the quasi-characteristic function of the two detectors and that, unknown, of 
the quantum system. This allows state reconstruction through the sequence: data collection,  
Fourier-transform, algebraic operation, inverse Fourier-transform. 
The strength of the measurement should be intermediate for the procedure to work. 
\end{abstract}
\maketitle
\emph{Introduction.} 
Quantum state tomography, i.e., the reconstruction of the unknown state of a
quantum system is a fundamental problem. Its formulation can be traced back to Pauli \cite{Pauli1933}, who asked whether 
a measurement of position on an ensemble of spinless systems prepared in a pure state, complemented by a measurement 
of momentum on a \emph{distinct} ensemble, would allow to reconstruct the wave function. 
This question has been answered in the negative \cite{Raymer1997,Corbett2006}. 
Pure states are represented by a wave-function, and they are exceptional, in the sense that they form a zero-measure subset of 
all possible states (which reflects in the great care one has to take in preparing a pure state).
The most general state is mixed and usually described by a density matrix, which is then the object to be reconstructed. 
However, other equivalent descriptions of a mixed state have revealed to be more useful or 
significant than the density matrix: the Wigner function \cite{Wigner1932}, 
the Husimi $Q$ function \cite{Husimi1940}, the Glauber-Sudarshan $P$ function
\cite{Glauber1963,Sudarshan1963}, the recently reintroduced Dirac function \cite{Chaturvedi2006}, and the almost forgotten 
Fourier-transform of Wigner function, which we refer to as the Moyal $M$ function \cite{Moyal1949}. 
Furthermore, parametric families including all of the above functions have 
also been introduced \cite{Cohen1966,Cahill1969}. 

Despite the quantum state being essential in describing a system, successful quantum state reconstruction, dubbed quantum state tomography 
because of an analogy with the germane procedure of medical tomography, is relatively recent \cite{Smithey1993}. 
The procedure relies on a proposal by Vogel and Risken \cite{Vogel1989}. 
For a recent review of continuous-variable quantum state tomography, the reader may refer to Ref.~\cite{Lvovsky2009}.

Recently, a remarkable experiment \cite{Lundeen2011} showed how it is possible to determine the unknown pure state of a 
one-dimensional quantum system by making a weak measurement of the $x$ variable
followed by a strong measurement of the conjugate variable $p$. 
A method, based on the Dirac function, allowing to lift the restriction to pure states was proposed 
recently \cite{Lundeen2012}.
Here, we make an alternative proposal, allowing the reconstruction of the Moyal $M$ function by a quick sequence of two measurements of 
conjugate variables. In order for the procedure to work, the strength of the measurement is not a fundamental issue, 
provided it is not too strong nor too weak. 
The equations at the basis of this proposal were reported, without derivation, in Ref. \cite{DiLorenzo2011}, and they are an exact result, not a perturbative expansion in the couplings.  
Furthermore, the procedure proposed herein requires a fixed setup. This is to be contrasted to homodyne quantum state tomography, where a different quadrature operator 
$x_\phi$ is measured for various values of $\phi$. The efficiency of the two procedures is otherwise comparable, as in the sequential measurements one has to evaluate a joint 
probability of two variables, while in the homodyne detection scheme a one-parameter family of single variable probabilities is estimated. 

\emph{Review of the measurement.} 
A linear measurement relies on a specific bilinear coupling between a quantum system and a probe. 
The interaction is assumed bilinear in the observable $\Hat{X}$ belonging to the system and in an 
observable $\Hat{\Phi}$ belonging to the probe: 
\begin{equation}
H_{int}=-\lambda\, \hbar\, g(t-t_0)  \Hat{\Phi} \Hat{X},
\end{equation}
where the function $g(t)$ is strongly peaked around $t=0$ and has unit integral. 
If the spectrum of $\Hat{X}$ is bounded, then $\lambda$ may include a scale such 
that the eigenvalues of $\Hat{X}$, indicated by $X$, are dimensionless and less than one\footnote{Unhatted variables refer to eigenvalues 
of the corresponding operators. Capital letters refer to the direct space, while lower-case letters refer to the conjugate space. E.g., $k$ is the 
conjugate of $X$, etc.}. 
A detector with a continuous unbounded output is considered, 
so that $\Hat{\Phi}$ has a conjugate operator $\Hat{J}$, satisfying $[\Hat{\Phi},\Hat{J}]=i$. 
Notice that $\lambda X$ and $J$ are homogeneous. 
The variable $J$ represents eigenvalues of $\Hat{J}$ and is the readout of the detector, carrying information about the system. 
If the probe is initially in a well defined state with vanishing variance in the readout variable $J$, 
i.e. its density matrix is $\rho(J,J')=\delta_{J,J'} \delta(J)$,  
then the measurement is an ideal strong one (we are indicating by $\delta_{a,b}$ the Kronecker delta and by $\delta(a)$ the Dirac delta). 
When this latter requirement is relaxed the measurement is a linear ideal (non-strong) one.  
More precisely, let $X_m$ be the typical spacing between the eigenvalues of $\Hat{X}$; 
the measurement is weak when the coupling constant satisfies  
$\lambda  X_m\ll \Delta$, with $\Delta^2$ the initial variance of the readout. 
One can distinguish two regimes: the weak incoherent measurement, 
when $\kappa \ll \lambda  X_m\ll \Delta$, with $\kappa$ the coherence scale of the detector (in the readout basis), and 
the weak coherent measurement,  when $\lambda  X_m\ll \kappa$. 
The former case bears little interest: because of the large variance, the readout of the detector in each individual 
trial is not necessarily $\lambda X$ and can lie well outside the spectrum, but 
after averaging over many trials, this effect washes out, even if one post-selects the system \cite{DiLorenzo2008,Cho2010}. 
The latter case was shown to produce a large average output after post-selection \cite{Aharonov1988,Duck1989}, and 
to allow a joint measurement of non-commuting observables with optimal noise in both outputs 
\cite{Arthurs1965,DiLorenzo2011}. 
It is interesting to note that a measurement of a continuous variable ($X_m=0$) is always a weak coherent measurement, 
and thus may show quantum coherence effects when followed by a post-selection. 
In the following, no specific assumption about $\lambda$ are made, in order to have general results. 
For simplicity an instantaneous interaction $g(t)=\delta(t)$ is considered.

In lack of any sufficient reason to believe otherwise, the probe and the system are assumed to be initially uncorrelated, 
so that their state immediately before the interaction is $\Hat{R}_-=\Hat{\rho}_S \otimes \Hat{\rho}_i$. 
We shall indicate by $\rho_i(J,J')=\langle J|\Hat{\rho}_i|J'\rangle$ [$\rho_S(X,X')=\langle X|\Hat{\rho}_S|X'\rangle$] 
the elements of the probe (system) density matrix in the $J$ ($X$) basis and 
by $\check{\rho}_i(\Phi,\Phi')=\langle \Phi|\Hat{\rho}_i|\Phi'\rangle$ [$\check{\rho}_S(K,K')=\langle K|\Hat{\rho}_S|K'\rangle$]
the elements in the $\Phi$  ($K$) basis.  
The system-probe state immediately after the interaction is
\begin{equation}\label{eq:finalstate}
R_{S,f}(X,X';J,J')= \rho_S(X,X') \rho_i(J-\lambda X, J'-\lambda X'),
\end{equation} 
where the equality $\exp{[-i J_0\Hat{\Phi}]}|J\rangle=|J-J_0\rangle$ was used. 
After tracing out the system, the final state of the probe following the interaction is 
\begin{equation}\label{eq:finalprobe}
\rho_f(J,J')=\int d\mu(X)\, \rho_S(X,X) \rho_i(J-\lambda X, J'-\lambda X),
\end{equation} 
with $\mu(X)$ the spectral function describing the 
distribution of eigenvalues (for a discrete spectrum, it is a combination of Dirac $\delta$-s, for a continuous one 
generally it is $d\mu(X)=dX$).
The probability distribution $\Pi_f$ of the readout $J$ is then 
\begin{equation}\label{eq:finalp}
\Pi_f(J)=\rho_f(J,J)=\int d\mu(X)\, \rho_S(X,X) \Pi_i(J-\lambda X), 
\end{equation} 
with $\Pi_i$ its initial distribution, 
and the corresponding characteristic function is 
\begin{equation}\label{eq:finalchi}
Z_f(\phi)\equiv\int dJ \, e^{i\phi J}\Pi_f(j)=Z_S(\lambda\phi) Z_i(\phi).
\end{equation} 
Equation~\eqref{eq:finalchi} reveals that the contribution of the detector to the cumulants 
(logarithmic derivatives of $Z_f$) are simply additive. In particular, if the detector is initially prepared 
in a Gaussian state, the cumulants of the output starting from the third and higher reflect faithfully the cumulants of the system, 
and only the second cumulant, which is the variance, includes a contribution from the detector, while the first one, the average, 
has a contribution from the probe only if this is biased, introducing a systematic error. 

\emph{Moyal quasi-characteristic function.}
The results will be specialized to a one-dimensional system, so that $X$ represents its coordinate, and 
$P=\hbar K$ its momentum. 
A useful transform of the density matrix was introduced in Ref.~\cite{Moyal1949}: the 
Moyal quasi-characteristic function which is but the Fourier transform of $W(K,X)$, 
the Wigner quasi-probability function \cite{Wigner1932,Moyal1949}, and is defined by 
\begin{align}
\nonumber
M(x,k)=&\!\! \int\!\! dX dK\,  e^{i k X+i x K}\, W(K,X)
\\
\nonumber
=&\!\! \int\!\! dX\,
 e^{i k X} \rho(X\!+\!\frac{x}{2},X\!-\!\frac{x}{2})\\
\label{eq:qchidef}
=& 
\!\! \int\!\! dK\,
 e^{i K x} \check{\rho}(K\!-\!\frac{k}{2},K\!+\!\frac{k}{2}) .
\end{align}
From the definition Eq.~\eqref{eq:qchidef} one realizes that $M(x,0)$ is the
characteristic function for the probability $\check{\Pi}_f(K)=\langle K|\Hat{\rho}|K\rangle$, 
and $M(0,k)$ the characteristic 
function for the probability ${\Pi_f}(X)=\langle X|\Hat{\rho}|X\rangle$. 
The generalization to a higher dimension is straightforward. 
The Moyal quasi-characteristic function uniquely determines the density matrix of a system, and \emph{vice versa}.
For composite systems, the marginal quasi-characteristic function of a subsytem is obtained 
by putting the coordinates of the remaining subsystems to zero in the total function. 
Furthermore, the Moyal function can be expressed as the average of the non-Hermitian Weyl operator
\begin{equation}
M(x,k)=\mathrm{Tr}\left\{\Hat{\rho}\, e^{ix\Hat{K}+ik\Hat{X}}\right\} ,
\end{equation}
where we used in the second line of Eq.~\eqref{eq:qchidef}
$|X-x/2\rangle=\exp[i\Hat{K} x/2] |X\rangle$, exploited the cyclic property of
the trace, and applied the Campbell-Baker-Hausdorff formula.

The usefulness of this transform shows, e.g., when considering the measurement illustrated in the previous 
section. The joint Moyal quasi-characteristic function for system and probe in terms of the initial ones is simply
\begin{equation}
M_{S,f}(x,k;\phi,j)=M_S(x,k+\lambda \phi) M_i(\phi,j-\lambda x).
\end{equation}
%%%

\emph{Results.}
A system interacting in rapid sequence with two probes, one coupling to $X$, the other to $K$, 
is  considered, so that the interaction term is 
\begin{equation}
H_{int}=-\hbar\left[\lambda_X  \delta(t+\varepsilon)  \Hat{\Phi}_X \Hat{X}+\lambda_K  \delta(t-\varepsilon)  \Hat{\Phi}_K \Hat{K}\right] .
\end{equation}
For $\varepsilon\to 0^-$ a measurement of $K$ is followed by a measurement of $X$, 
and \emph{vice versa} for $\varepsilon\to 0^+$. For $\varepsilon= 0$ the measurement is a joint measurement 
\emph{\`{a} la} Arthurs and Kelly \cite{Arthurs1965}, 
that we shall not study in detail here, and for which we refer the reader to our previous 
paper \cite{DiLorenzo2011}. 
The initial state is assumed to be $\Hat{R}_-=\Hat{\rho}_S\otimes \Hat{\rho}_i$, 
with $\Hat{\rho}_i$ the density matrix of the two probes immediately before the first interaction. 
The possibility that the probes are initially in a correlated state is accounted for. 
The quasi-characteristic function for the system and the two probes, after some straightforward calculations \cite{suppl}, 
turns out to be 
\begin{align}\label{eq:maintotal}
\nonumber
M_{S,f}(x,k;\phi,j) =& M_i\left[\phi,j+\Lambda\left(2\alpha_{0} s +\alpha_{\varepsilon} \Lambda\phi\right)\right]
 \\
&\times M_S(s+\Lambda \phi),
\end{align}
where for conciseness the two probes's coordinates were arranged in two column vectors 
$\phi=(\phi_K,\phi_X)$, $j=(j_K,j_X)$, $s=(x,k)$ represents the symplectic coordinates,  
we introduced the $2\times 2$ matrix $\Lambda=\mathrm{diag}(\lambda_K,\lambda_X)$, 
$\varepsilon\in\{+,-,0\}$, and 
\begin{align}
\alpha_{+}&=\begin{pmatrix}0&0\\ 1&0\end{pmatrix}\ ,\  
\alpha_{-}=\begin{pmatrix}0&-1\\ 0&0\end{pmatrix}\ ,\  
\alpha_{0}=\frac{\alpha_{+}+\alpha_{-}}{2}.
\end{align}
The case $\varepsilon=+$ corresponds to a measurement of $X$ followed by one of $K$, 
$\varepsilon=-$ to the opposite order, and $\varepsilon=0$ to the joint measurement. 
Then, the (proper) characteristic function, i.e. the Fourier transform of the
joint probability of observing $J=(J_K,J_X)$ as the output, is obtained 
by tracing out the system ($k=0,x=0$), and by putting $j=0$, so that 
the experimentally accessible quantity
\begin{equation}\label{eq:expz}
Z_f(\phi)=\int dJ_K dJ_X e^{iJ\cdot \phi} \Pi_f(J_K,J_X)
\end{equation}
is expressed in terms of the product of the target $M_S$ and the known $M_i$
\begin{align}
\label{eq:main}
Z_f(\phi)=&
M_{S}(\Lambda \phi) 
M_{i}(\phi,\Lambda\alpha_{\varepsilon}\Lambda\phi)  .
\end{align}
Equations \eqref{eq:maintotal} and \eqref{eq:main} are the main results of this paper. 
Since performing a joint measurement may be more difficult than a sequential one, 
we concentrate on the cases $\varepsilon=\pm$, but in principle the joint measurement of 
position and momentum allows as well the reconstruction of the initial state.

If the probes are initially in a known state, then the unknown initial state of the system can be 
evinced after dividing the left hand side of Eq.~\eqref{eq:main} by 
$M_{i}(\phi,\alpha_{\varepsilon})$. 
This can be done as far as 
$M_{i}(\phi,\alpha_{\varepsilon})\neq 0$. 
(A partial reconstruction of the state of the system may be satisfactory, however, if, e.g., 
the function $M$ is known on a dense set over $\mathbb{R}^2$ or everywhere but 
on a zero-measure set. See Ref.~\cite{Kiukas2012} for a detailed study.)
The density matrix and the Wigner function are given respectively by 
\begin{align}
\label{eq:main2}
\rho\left(X\!+\!\frac{x}{2},X\!-\!\frac{x}{2}\right)=&\int \frac{dk}{2\pi}  
\frac{e^{-i Xk} Z_f(V s)}{M_{i}(V s,\Lambda\alpha_{\varepsilon} s)},
\\
\label{eq:main3}
\check{\rho}\left(K\!-\!\frac{k}{2},K\!+\!\frac{k}{2}\right)=&\int \frac{dx}{2\pi}  
\frac{e^{-i K x} Z_f(Vs)}{M_{i}(V s,\Lambda\alpha_{\varepsilon} s)},\\
\label{eq:main4}
W(K,X)=&\int \frac{dk dx}{(2\pi)^2} 
\frac{e^{-i (K x+X k)} Z_f(Vs)}{M_{i}(V s,\Lambda\alpha_{\varepsilon}s)},
\end{align}
with $V=\Lambda^{-1}=\mathrm{diag}(\lambda_K^{-1},\lambda_X^{-1})$.

%
%%  
%%%
For an ideal strong measurement, the initial pointer density matrix would be 
$\rho_i(J,J')=\delta_{J_K,J'_K} \delta_{J_X,J'_X}\delta(J_K)\delta(J_X)$, and hence 
$M_{i}(\phi,j)=\delta_{j_K,0} \delta_{j_X,0}$, 
so that the procedure would not work for $\varepsilon=+,\ \phi_K\neq 0$, nor for $\varepsilon=-,\ \phi_X\neq 0$, nor for $\varepsilon=0,\ \phi_K\neq 0 \ \text{or } \phi_X\neq 0$. 
It is therefore desirable not to work in the strong regime, but at the same time one does not need 
to keep the measurement in the weak regime. 
For instance, the probes could be prepared in the mixed gaussian state
\begin{equation}
\rho_i(J,J')=\frac{\exp{\left\{-\sum_a [\bar{J}_{a}^2/2\Delta_a^2+j_a^2/2\kappa_a^2]\right\}}
}{2\pi \Delta_K \Delta_X} ,
\end{equation}
with $\bar{J}_{a}=(J_a+J'_a)/2$, ${j_a}=(J_a-J'_a)$,  
and $\kappa_a$ representing the coherence scale (that satisfies $\kappa_a=1/\tilde{\Delta}_a$, 
where $\tilde{\Delta}_a$ is the spread of the conjugate variable $\Phi_a$, 
so that by the uncertainty principle $\kappa_a\le 2\Delta_a$). 
Then 
\begin{equation}\label{eq:gauss}
M_i(\phi,j)=
\exp{\left\{-\sum_a \left[\phi_{a}^2\Delta_a^2/2+j_a^2/2\kappa_a^2\right]\right\}}.
\end{equation}
The ideal strong case is obtained for $\kappa_a\to 0$, $\Delta_a\to 0$.  
In the ideal weak coherent regime, $\kappa_a\to \infty$, $\Delta_a\to \infty$, the state is 
$M_i(\phi,j)\to \delta_{\phi_K,0}\delta_{\phi_X,0}$ (which 
is but the strong regime for when the $\Phi$-variables are used as a readout), hence 
Eq.~\eqref{eq:main} can not determine  
$M_S(\lambda_K\phi_K,\lambda_X\phi_X)$ for $\phi_K,\phi_X\neq 0$.  
Thus the procedure suggested here works with intermediate measurement strength. 

In particular, when the detector is prepared in the state given by Eq.~\eqref{eq:gauss}, 
the Wigner function in terms of the characteristic function of Eq.~\eqref{eq:expz} is 
\begin{widetext}
\begin{align}
\nonumber
W(K,X)=\int \frac{dk dx}{(2\pi)^2} Z_f(Vs)
\exp\Biggl\{-i (K x+X k)&+ \frac{1}{2}\left[\left(\frac{\Delta_K}{\lambda_K}\right)^2
+\left(\frac{(1+\varepsilon)\lambda_X}{2\kappa_X}\right)^2\right]x^2 \\
\label{eq:special}
&+ 
\frac{1}{2}\left[\left(\frac{\Delta_X}{\lambda_X}\right)^2+\left(\frac{(1-\varepsilon)\lambda_K}{2\kappa_K}\right)^2\right] k^2\Biggr\}
.
\end{align}
\end{widetext}
After substituting Eq.~\eqref{eq:expz} into Eq.~\eqref{eq:special}, one should be careful not to 
exchange recklessly the order of integration, or an artificial divergence appears. 
Furthermore, for $\varepsilon=0$ and $\kappa_a=2\Delta_a$, $\Delta_X\Delta_K=\lambda_X\lambda_K/4$, 
Eqs.~\eqref{eq:special} and  \eqref{eq:expz} give a relation between $W(K,X)$ and $\Pi_f(J)$ that is the 
formula relating the Wigner and the $Q$ function, so that, as is well known \cite{Braunstein1991}, 
the joint measurement of position and momentum provides directly the $Q$ function, provided the detectors are properly 
prepared. By contrast, the general procedure proposed here for measuring the Moyal $M$ function allows much more 
flexibility in terms of detectors preparation.

Finally, an alternative application of Eq.~\eqref{eq:main} could consist in   
obtaining the quasi-characteristic function of the system for two fixed values of 
$\phi_K,\phi_X$ and then vary the coupling strengths $\lambda_K,\lambda_X$ 
by keeping $\lambda_K \lambda_X$ fixed 
in order to reconstruct the state. 

%%%
%%
%
\emph{Estimates.}
Let us summarize the steps needed to proceed to the quantum state tomography with the method proposed here, which we may call quantum sequential tomography (QST), 
while comparing them to the analogous steps done in quantum homodyne tomography (QHT). 

In QST, first, a joint probability $\Pi_f(J_K,J_X)$ is measured by observing the $J$ variables of the probe. 
Experimentally, one should divide the $J_K,J_X$ plane in a sufficient number of small bins, each having an area $\Delta J_X \Delta J_K$ determined by the precision of the measurement. 
The measurement is repeated a large number of times $N$, so that the resulting histogram approximates the true probability within the precision of the probes. 
Analogously, in QHT, the conditional probability $pr(x_\phi|\phi)$ is measured for different values of $\phi\in[0,\pi]$. 
The binning is in the $x,\phi$ strip, and it is determined by the precision with which $x_\phi$ can be measured and $\phi$ controlled. 
Then, in QST, the Moyal function $M_S(x,k)$ is obtained by making a double integral, precisely a Fourier transform of the observed probability, 
which can be done efficiently fast with the Fast-Fourier Transform algorithm, and by dividing finally by the known state of the probes, which is computationally trivial. 
In QHT, the Wigner function is obtained by making  an inverse Radon transform of $pr(x_\phi|\phi)$, which consists as well in a double integral. 
Finally, the density matrix can be obtained by making a single Fourier transform of the Moyal characteristic function in QST, and of the Wigner function in QHT. 
A further Fourier transform is needed to obtain the Wigner function in QST and the Moyal function in QHT.
Thus we can say that QST is more efficient of QHT for the determination of the Moyal function, at least as performant as QHT for the density matrix and less efficient for 
obtaining the Wigner function. 

Finally, we estimate the relative uncertainty in the Moyal function, $M_S(s)=Z_f(V s) M_i^{-1}(V s,\Lambda\alpha_\varepsilon  s)$. 
As $M_i$ is fixed, it carries an uncertainty $\delta M_i(\phi,j)$ that depends on how the state of the probes was determined.   
The finite sampling introduces a statistical error in $Z_f(\phi)$, that can be estimated according to standard statistical analysis as $\delta Z^2 = [1 -|Z_f(\phi)|^2]/N$. 
There is also a numerical error introduced by the integration, $\eta_{\mathrm{num}}$. Notice how these uncertainties are present also in QHT. 
In conclusion, in QST, the relative error can be estimated as 
\begin{align}
\frac{\delta |M_S(s)|^2}{|M_S(s)|^2} \sim& \frac{1-|Z_f(V s)|^2}{N |Z_f(V s)|^2} +\frac{\delta |M_i(V s,\Lambda\alpha_\varepsilon  s)|^2}{|M_i(V s,\Lambda\alpha_\varepsilon  s)|^2}
\nonumber \\
&\qquad \qquad \qquad \qquad + \eta^2_{\mathrm{num}}(Vs).
\end{align}

\emph{Conclusions.}
A procedure for determining an unknown quantum state was proposed. 
Two measurements of conjugate variables are made in quick sequence, the joint probability 
is estimated from the collected data, it is Fourier-transformed to give the characteristic function, 
then divided by the quasi-characteristic function of the probes appearing in the right hand side 
of Eq.~\eqref{eq:main}. 
This yields the Moyal quasi-characteristic function of the system. 
The density matrix is obtained by Fourier-transforming the latter function. 
On the other hand, it is sufficient to determine the Moyal function only in a
neighborhood of $x=0,k=0$ in order to estimate the cumulants. 
Furthermore, the method proposed has the advantage of requiring one fixed setup and does not require a sharp measurement of either 
position or momentum, rather it thrives over the unsharpness of the measurement. 

In perspective, it would be interesting to extend the results to finite-dimensional Hilbert spaces, for which there is a wide interest (see, e.g., the recent Ref.~\cite{Kalev2012}), 
especially in the light of a recent generalization \cite{Carmeli2011} of the concept of conjugate variables. 

I acknowledge stimulating discussions with J. Lundeen. 
This work was performed as part of the Brazilian Instituto Nacional de Ci\^{e}ncia e
Tecnologia para a Informa\c{c}\~{a}o Qu\^{a}ntica (INCT--IQ) and 
it was supported by Funda\c{c}\~{a}o de Amparo \`{a} Pesquisa do 
Estado de Minas Gerais through Process No. APQ-02804-10.

%\bibliography{../weakmeasbiblio}
\begin{widetext}
\section{Supplemental Material:\\ Derivation of the main result.}	
The case of a measurement of momentum followed by one of position is considered for definiteness. 
The coupling constants are temporarily absorbed in the rescaled variables 
$\Phi_a\to \lambda_a \Phi_a$, $J_a\to J_a/\lambda_a$, $a=K,X$.  
The joint density matrix after the interaction is 
\begin{equation}
\Hat{R}_{S,f}=e^{i\Hat{\Phi}_X\Hat{X}}e^{i\Hat{\Phi}_K \Hat{K}}
\Hat{R}_-  
e^{-i\Hat{\Phi}_K \Hat{K}}e^{-i\Hat{\Phi}_X\Hat{X}},
\end{equation}
with $\Hat{R}_-=\Hat{\rho}_S \otimes \Hat{\rho}_i$ the initial joint density matrix.
The most convenient representation is in terms of $|\Phi\rangle$ for the probes and $|X\rangle$for the system 
\begin{align}
&{R}_{S,f}(\Phi,X,\Phi',X')
= e^{i{\Phi}_X X}
\rho_S(X+\Phi_K,X'+\Phi'_K) \rho_i(\Phi,\Phi')  
e^{-i\Phi'_X X'},
\end{align}
where $\exp[-i a \Hat{K}]|X\rangle=|X+a\rangle$ was used. 
Multiplication by $\exp[i k\bar{X}]$ and by $\exp[i j\cdot\bar{\Phi}]$, 
with $\bar{X}=(X+X')/2$ and $\bar{\Phi}=(\Phi+\Phi')/2$, followed 
by integration over $\bar{X}$ and $\bar{\Phi}$ yields 
\begin{align}
\nonumber
{M}_{S,f}(x,k;\phi,j)
=&
\int d\bar{\Phi}d\bar{X}\,
e^{i[x{\bar{\Phi}}_X+j\cdot\bar{\Phi}+(k+\phi_X)\bar{X}]} \rho_i(\bar{\Phi}+\frac{\phi}{2},\bar{\Phi}-\frac{\phi}{2})  
\rho_S(\bar{X}+\bar{\Phi}_K+\frac{x+\phi_K}{2},\bar{X}+\bar{\Phi}_K-\frac{x+\phi_K}{2})
\\
\nonumber
=&
\int d\bar{\Phi}d{Y}\,
e^{i[x{\bar{\Phi}}_X+j\cdot\bar{\Phi}+(k+\phi_X)(Y-\bar{\Phi}_K)]}
\rho_i(\bar{\Phi}+\frac{\phi}{2},\bar{\Phi}-\frac{\phi}{2})  
\rho_S(Y\!+\!\frac{x+\phi_K}{2},Y\!-\!\frac{x+\phi_K}{2}) \\
\nonumber
=&
\int d\bar{\Phi}\,
e^{i[x{\bar{\Phi}}_X+j\cdot\bar{\Phi}-(k+\phi_X)\bar{\Phi}_K]} \rho_i(\bar{\Phi}+\frac{\phi}{2},\bar{\Phi}-\frac{\phi}{2})  
M_S(x+\phi_K,k+\phi_X)
\\
=&\ 
M_i(\phi,j+\gamma)
M_S(x+\phi_K,k+\phi_X)
\end{align}
with $\gamma=(-k-\phi_X,x)$. 
The identity 
$\Phi_X X-\Phi'_X X' = \bar{X} \phi_X +\bar{\Phi}_X x$ 
was used, and the change of variable 
$Y=\bar{X}+\bar{\Phi}_K$ was made. 
\end{widetext}

\end{document}